\documentclass[twocolumn,prl,twocolumn,showpacs]{revtex4}
\usepackage{graphicx}
\usepackage{dcolumn}
\usepackage{bm}
\usepackage{epstopdf}

\begin{document}

\title{\bf  Calculations of spin induced transport in ferromagnets}

\author{M. S. Bahramy$^1$}
\email{saeed@imr.edu}
\author{ P. Murugan$^1$}
\author{G. P. Das$^2$} 
\author {Y. Kawazoe$^1$ }
\affiliation{$^1$Institute for Materials Research, Tohoku University, Aoba-ku, Sendai 980-8577, Japan \\
$^2$Indian Association for Cultivation of Science, Jadavpur, Kolkata 700032, India}

\date{\today}

\begin{abstract}
Based on first-principles density functional calculations, a general approach for determining and analyzing the degree of spin polarization ($P$) in ferromagnets is presented. The approach employs the so-called tetrahedron method to evaluate the Fermi surface integrations of $P$ in both ballistic and diffusive regimes. The validity of the method is examined by comparing the calculated $P$ values for Fe and Ni with the experiment. The method is shown to yield highly accurate results with minimal computational effort. Within our approach, it is also possible to systematically analyze the contributions of various types of electronic states to the spin induced transport. As a case study, the transport properties of the soft-ferromagnet CeMnNi$_4$ are investigated in order to explain the origin of the existing difference between the experimental and theoretical values of $P$ in this intermetallic compound.

\end{abstract}

\maketitle

\section{Introduction}

The current upsurge of interest in metal-based as well as semiconductor-based spintronics materials mainly originates from their increasing applications in the giant magnetoresistance (GMR) as well as the tunneling magnetoresistance (TMR) based memory devices~\cite{wolf2001, moodera1995}.
Common to all these devices is the requirement of controlled transfer or injection of spin-polarized current from a ferromagnet into a normal metal~\cite{katine2000} or a semiconductor~\cite{jaffres2002}.  For such a spin transport to be effective, it is mandatory to maintain a relative imbalance between spin-up and spin-down electrons that are available at Fermi level, $\varepsilon_F$. By definition, degree of spin polarization, $P$, is described as the extent to which the transport current is spin polarized. 
Depending on the degree and asymmetry of spin-splitting, different values of $P$, ranging from $\sim 40\%$ for a conventional ferromagnetic metal such as Fe to  $\sim 100\%$ for the so-called half-metallic systems ~\cite{groot1983} have been reported. 
In the latter, all the transport current is carried by  either spin-up or spin-down electrons.
Heusler alloys~\cite{ayuela1999}, chalcopyrites~\cite{medvedkin2000} are the best examples of half-metallic ferromagnets  whose
magnetic behaviour as well as electronic structure have been extensively studied both experimentally and theoretically. 
 
 In practice, the spectroscopic techniques, such as spin-polarized photoemission~\cite{johnson1997},  point contact Andreev reflection (PCAR)~\cite{soulen1998} and tunnel junction (TJ)~\cite{meservey1994}, are widely used to probe $P$ at $\varepsilon_F$ of a ferromagnet. However, except for the case of half metallic systems, different experiments may measure different values of $P$ for the same candidate material~\cite{nadgorny2001}. 
 The reason is that, in a realistic experiment, the observed tunneling current is dependent on both the mean free path of electrons in the ferromagnet, $l$, and the characteristic size of the contact $d$.
If $d$ is smaller than the electron mean free path $l$, the electrons
flow through the contact ballistically. In the opposite case, when $d\gg l$, they perform a diffusive motion. Consequently, the obtained values of $P$, in each regime, can be considerably different. In this connection, it is important to compare the measurements with the appropriate calculations, mimicking the experiment in question. 

Within the classical Bloch-Boltzmann theory~\cite{allen1978},  $P$  can be defined as $(J_\uparrow-J_\downarrow)/(J_\uparrow+J_\downarrow)$ where $J_{\uparrow (\downarrow)}$ are the spin-dependent current densities, passing through the contact. Neglecting the state-dependent transmittance of the barrier at the point of contact, Mazin~\cite{mazin1999} has formulated a more practical definition for $P$  as, 
\begin{equation}
P_{n}=\frac{\left\langle N\upsilon^n_F\right\rangle_\uparrow-\left\langle N\upsilon^n_F\right\rangle_\downarrow}{\left\langle N\upsilon^n_F\right\rangle_\uparrow+\left\langle N\upsilon^n_F\right\rangle_\downarrow}
\label{mazin}
\end{equation}
where, $N$ and $\upsilon_F$ are the density of states (DOS) at $\varepsilon_F$ and Fermi velocity of electrons with spin $\sigma$ ($\uparrow$ or $\downarrow$), respectively. Interestingly, all the quantities in eq.~\ref{mazin} are related to the band structure of the ferromagnet.  
Furthermore, this definition allows direct comparison between different experiments and theory. In the simplest case of $n=0$, the spin polarization  calculated from eq.~\ref{mazin} ($P_0$) is in accordance with that measured by spin-resolved photoemission techniques. 
Correspondingly, it can be shown that, the transport experiments, such as PCAR and TJ, measure a higher order of $P_n$ ($n=1$ or $2$) which includes the Fermi velocity (see eq.~\ref{mazin}). In the ballistic regime ($d\ll l$), following Sharvin's approach ~\cite{sharvin1965} or Landauer-B\"utikker formalism~\cite{schep1998}, one can easily prove that the conductance of the contact $G_\sigma$ and, consequently $J_\sigma$ are proportional to  $\left\langle N\upsilon_F\right\rangle_\sigma$ ~\cite{mazin1999,mazin2001,schep1998}. 
 Thus, the experimentally observed  $P$ in the ballistic regime is expected to be equal with $P_1$. On the other hand, using Bloch-Boltzmann equation, one can prove that, $G_\sigma$ and, consequently $J_\sigma$ in the diffusive regime ($d\gg l$) are proportional to  $\left\langle N\upsilon_F^2\right\rangle_\sigma\tau_\sigma$, where $\tau$ is the relaxation time~\cite{mazin1999,mazin2001}. Hence, assuming the same $\tau_\sigma$ in both spin channels, the spin polarization observed by transport techniques in diffusive regime, is expected to be in accordance with $P_2$. 

Computationally, it is rather difficult to calculate $\left\langle N\upsilon^n_F\right\rangle_\sigma$ for $n>0$. The problem is due to the presence of gradient term $v_F$=$\nabla_{\bf k}\varepsilon$, which makes calculations very sensitive to the number of ${ k-}$points, used for sampling the Brillouin zone. Consequently, in most of the theoretical works, it is preferred to compare experimental data with the calculated $P_0$. However, such a comparison may lead to a misleading conclusion in describing the transport properties of the material under consideration. In this work, we present a simple and general approach, based on the so called tetrahedron method~\cite{jepsen1971,lehmann1972}, for evaluating $P_n$ in both ballistic and diffusive limits with high accuracy. First, the details of the method will be explained and then the accuracy of $P_n$ values for Fe and Ni will be examined by comparing our results with the experiment. 
Finally, as a case study, the transport properties of the recently discovered soft ferromagnet CeMnNi$_4$ are investigated, in order to find out the underlying reasons for the pronounced difference between the experimental~\cite{singh2006} and theoretical~\cite{murugan2006,voloshina2006,mazin2006} values of $P$ in this particular intermetallic compound.

\section{methodology}
As a general definition, the expectation term of eq.~\ref{mazin}, $\left\langle{N\upsilon^n}\right\rangle_\sigma$, can be expressed as~\cite{mazin1999},
\begin{eqnarray}
\left\langle{N\upsilon^n}\right\rangle_\sigma &=& \frac{1}{(2\pi)^3}\sum_\lambda\int\upsilon^n_{{\bf k}\lambda\sigma}\delta(\varepsilon_{{\bf k}\lambda\sigma}-\varepsilon_F)d^3k \nonumber \\
                                            &=& \frac{1}{(2\pi)^3}\sum_\lambda\int \upsilon^{n-1}_{{\bf k}\lambda\sigma}dS_F,	
\label{ballusive}
\end{eqnarray} 
where, $\varepsilon_{{\bf k}\lambda\sigma}$ is the energy of an electron in the band $\lambda$  with spin $\sigma$ and the wave vector ${\bf k}$. At this point, to avoid any confusion, we neglect the subscripts $\lambda$ and $\sigma$ and simplify the integration part of eq.\ref{ballusive} as,
\begin{equation}
 \jmath(\varepsilon_F)=\int \frac{dS_F}{\left|\nabla_{\bf k}\varepsilon\right|}A({\bf k})
 \label{tetint}	
\end{equation}
where, A({\bf k}) corresponds to $\upsilon^n_F=\left|\nabla_{\bf k}\varepsilon\right|^n$. 

For evaluation of the integral $\jmath(\varepsilon_F)$ over Fermi surface, $S_F$, we consider the following scheme: First we define an equispaced grid in reciprocal space on which the lattice vectors of submesh are obtained by dividing a set of primitive reciprocal vectors by the integers $n_1$, $n_2$ and $n_3$. 
Then, utilizing the Monkhorst-Pack method~\cite{monkhorst1976} a set of  irreducible $k-$points from $\left\{ n_1,n_2,n_3 \right\}$ is chosen for which the eigenvalues of energy, $\varepsilon_{\bf k}$, are calculated. For the $k-$points, which are not in the irreducible part of reciprocal space, symmetry can be used to determine their corresponding $\varepsilon_{\bf k}$ in the irreducible zone.    

\begin{figure}[t]
  \begin{center}
\rotatebox{-0}{\includegraphics[width=3.2 in]{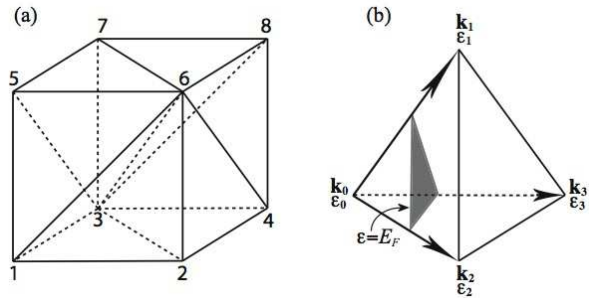}} 
  \end{center}
  \caption{Schematic illustration of (a) a subcell and its division into six tetrahedrons of same volume and (b) one tetrahedron spanned out by the vectors $\bf{k}_0$, $\bf{k}_1$, $\bf{k}_2$, and $\bf{k}_3$ with the corresponding eigenvalues $\varepsilon_0$, $\varepsilon_1$, $\varepsilon_2$ and $\varepsilon_3$, respectively. The hatched plane is the plane with $\varepsilon=\varepsilon_F$. }
  \label{fig:tetrahedron}
\end{figure}

 Next, each subcell is divided into six tetrahedra of same volume. As illustrated in Fig~\ref{fig:tetrahedron}, we choose one main diagonal of a subcell as common edge of all six tetrahedra. In order to minimize interpolation distance, the shortest main diagonal is chosen.
 Additionally, each tetrahedron is defined by its four corners, one of which is, without any restriction of generality, at the origin, $\bf{k}_0={\bf 0}$, and the remaining three are at $\bf{k}_i$ $(i=1-3)$. With some rearrangement, the four $\bf{k}_i$ are ordered so that their calculated eigenvalues $\varepsilon_i$  obey the  inequality  $\varepsilon_0\leq\varepsilon_1\leq\varepsilon_2\leq\varepsilon_3$. 

Finally, within each tetrahedron, $\varepsilon=\varepsilon_F$ is interpolated by a linear function,
\begin{equation}
	\varepsilon=\varepsilon_0+{\bf b}\cdot{\bf k},
	\label{interp}
\end{equation}  
 where, ${\bf b}$  can  be described in terms of $\varepsilon_i$ and ${\bf k}_i$. In this order, the triple ${\bf r}_i$, contragradient to ${\bf k}_i$ are defined as,
\begin{equation}
{\bf r}_1=\frac{{\bf k}_2\times{\bf k}_3}{V};~~{\bf r}_2=\frac{{\bf k}_3\times{\bf k}_1}{V};~~ {\bf r}_3=\frac{{\bf k}_1\times{\bf k}_2}{V}
	\label{ri}
\end{equation}
where $V$ is the volume of subcell, $V=\left|{\bf k}_1\cdot({\bf k}_2\times{\bf k}_3)\right|$, and ${\bf r}_i$ satisfy the constraint, ${\bf r}_i\cdot{\bf k}_j=\delta_{ij}$. Using eq.~\ref{ri}, ${\bf b}$ can be expressed as,
\begin{equation}
{\bf b}=\sum_i(\varepsilon-\varepsilon_i){\bf r}_i
	\label{bvector}
\end{equation}

Back to integral~\ref{tetint},   $A({\bf k})$ can be written as a function of ${\bf b}$, $A({\bf k})=\left|{\bf b}\right|^n$. Accordingly, $\jmath(\varepsilon_F)$ over one tetrahedron becomes,
\begin{equation}
\imath_{\bf T}(\varepsilon_F)=\int_{\bf T}{dS_F}\left|{\bf b}\right|^{n-1}
	\label{inti}
\end{equation}
and the integration reduces to the following simple form,
\begin{equation}
\imath_{\bf T}(\varepsilon_F)=\xi(\varepsilon_F)\left|{\bf b}\right|^{n-1}
	\label{i0}
\end{equation}
where $\xi(\varepsilon_F)$ is the cross section of the plane $\varepsilon=\varepsilon_F$ with the tetrahedron (see Fig~\ref{fig:tetrahedron}). Inserting eq.~\ref{ri} into eq.~\ref{bvector} and using the shorthand notation $\varepsilon_{ij}$ for $\varepsilon_{i}-\varepsilon_{j}$, one can easily prove that $\imath_{\bf T}(\varepsilon_F)$ becomes,
\begin{equation}
\imath_{\bf T}(\varepsilon_F)=\frac{V}{2}\frac{(\varepsilon_{F}-\varepsilon_{0})^2}{\varepsilon_{10}\varepsilon_{20}\varepsilon_{30}} \left|{\bf b}\right|^{n}
	\label{line1}
\end{equation} 
if $\varepsilon_0\leq\varepsilon_F\leq\varepsilon_1$,
\begin{equation}
\imath_{\bf T}(\varepsilon_F)=\frac{V}{2} \left[\frac{(\varepsilon_{F}-\varepsilon_{0})^2}{\varepsilon_{10}\varepsilon_{20}\varepsilon_{30}} -\frac{(\varepsilon_{F}-\varepsilon_{1})^2}{\varepsilon_{10}\varepsilon_{21}\varepsilon_{31}} \right]\left|{\bf b}\right|^{n}
	\label{line2}
\end{equation} 
if $\varepsilon_1\leq\varepsilon_F\leq\varepsilon_2$, and
\begin{equation}
\imath_{\bf T}(\varepsilon_F)=\frac{V}{2}\frac{(\varepsilon_{3}-\varepsilon_{F})^2}{\varepsilon_{30}\varepsilon_{31}\varepsilon_{32}} \left|{\bf b}\right|^{n}
	\label{line3}
\end{equation} 
if $\varepsilon_2\leq\varepsilon_F\leq\varepsilon_3$.

Here, it is important to point out that, within the improved tetrahedron method, proposed by Bl\"ochl et al.~\cite{blochl1994}, the linear interpolation in eq.~\ref{line2} can be modified as,
\begin{eqnarray}
\imath_{\bf T}(\varepsilon_F)&=&\frac{V}{2}\frac{\left|{\bf b}\right|^{n}}{\varepsilon_{20}\varepsilon_{30}}\times \nonumber \\ &&\left[{\varepsilon_{10}+2(\varepsilon_{F}-\varepsilon_{1})-\frac{(\varepsilon_{20}+\varepsilon_{31})(\varepsilon_{F} -\varepsilon_{1})^2} {\varepsilon_{21}\varepsilon_{31}}}\right]\nonumber\\
	\label{line2b}
\end{eqnarray}
Accordingly, throughout this work eq.~\ref{line2b} has been considered instead.

At the last step, summation over $\imath_{\bf T}(\varepsilon_F)$ from contributing tetrahedra results in the evaluation of integral $\jmath(\varepsilon_F)$. Here, it is crucial to point out that, the dependence of $\imath_{\bf T}(\varepsilon_F)$ on the geometry of tetrahedrons for $n>0$, implies that the summation  should be carried out  over all tetrahedra in the reciprocal space, otherwise a misleading error will occur in the evaluation of the integrals.

In the following, we apply this prescription to calculate $P_n$ for a number of ferromagnetic materials, using their electronic band structure obtained from the first-principles calculations. While our method is quite general and applicable to any system, we first present some benchmark calculations of $P_0$, $P_1$ and $P_2$ for bcc-Fe (a$=2.866 $ \AA) and fcc-Ni (a$=3.524 $ \AA), and compare the results with the experimental data, available in the literature~\cite{meservey1994, soulen1998}. Then, we compare the $P_n$ values for CeMnNi$_4$  obtained from the current method with those reported in previous theoretical investigations~\cite{voloshina2006,mazin2006} as well as experiment~\cite{singh2006}. Here, we focus on the transport spin-polarization of cubic CeMnNi$_4$ (a$=6.987 $ \AA) containing 4 formula units i.e. 4 Ce, 4 Mn and 16 Ni atoms.  For the generated irreducible ${\it k-}$points (see discussion above) the eigenvalues, $\varepsilon_{{\bf k}\lambda\sigma}$, of all the available energy bands, $\lambda$, at $\varepsilon_F$ are calculated in both spin channel $\sigma$ and used for evaluating $P_n$.  All the first-principles calculations are carried out using the local spin density approximation (LSDA) method with projector-augmented wave (PAW) potential, as implemented in the VASP code~\cite{vasp}. Details of the electronic structure calculations have been published elsewhere~\cite{murugan2006}.

\section{Results and Discussion} 
Table~\ref{tab:spinp} summarizes the $P$ values for Fe, Ni, and CeMnNi$_4$, as obtained from our calculations ($P_n$, $n=0-2$), and as obtained from PCAR and TJ measurements ($P_C$ and $P_T$, respectively). Here, we first concentrate on the transport properties of Fe and Ni. For both bulk ferromagnetic metals, the table indicates an excellent agreement between $P_1$ ($P_2$) and $P_C$ ($P_T$) values. Such an agreement indicates that, the experimental measurements of $P_C$ and $P_T$ for both metals have been carried out in ballistic and diffusive regimes, respectively. Fortunately, the details of the corresponding experiments explicitly confirm this statement~\cite{meservey1994,soulen1998}.
\begin{table}[t]
\caption{\label{tab:spinp} Comparison of $P$ values (in \%) for Fe, Ni, and CeMnNi$_4$, as obtained from our calculations ($P_n$ ($n=0-2$)), and as obtained from PCAR and TJ measurements ($P_C$ and $P_T$, respectively). The table also includes the calculated averaged Fermi velocities $\left\langle{\upsilon}_F\right\rangle$ in both spin channels (in ${10^7}$cm/s). }
\begin{ruledtabular}
\begin{tabular}{lccccccc}
&\multicolumn{5}{c}{Present work}&\multicolumn{2}{c}{Experiment}\\ 
 \cline{2-6}  \cline{7-8}
 \\ [0.5pt]
Structure & $P_0$ & $P_1$ & $P_2$ & $\left\langle{\upsilon}_F\right\rangle_\uparrow$ & $\left\langle{\upsilon}_F\right\rangle_\downarrow$ & $P_C$\footnotemark[1] & $P_T$\footnotemark[2]\\ [2pt]
\hline
\\
Fe        & $  54$ & $ 42$ & $ 37$ & $ 3.1$ & $4.4$ & $43$     &$   40$   \\[2pt]
Ni        & $ -80$ & $-47$ & $  15$ & $ 5.5$ & $1.6$ & $\pm46.5$&$   23$   \\[2pt]
CeMnNi$_4$& $ -22$ & $ -5$ & $ 10$ & $ 2.2$ & $1.5$ & $\pm66$\footnotemark[3]&$---$
\end{tabular}
\end{ruledtabular}
\footnotetext[1]{from Ref.~\onlinecite{soulen1998}}
\footnotetext[2]{from Ref.~\onlinecite{meservey1994}}
\footnotetext[3]{from Ref. ~\onlinecite{voloshina2006}}
\end{table}

Furthermore, from table~\ref{tab:spinp}, one immediately notices two different trend of spin polarizations between Fe and Ni. In the former, the differences between $P_0$, $P_1$ and $P_2$ values are moderate while in the latter, such  differences are so considerable that leads to a change in the sign of $P_2$. 
The reason can be explained  in terms of the electronic band structure of transition metals at the Fermi level.  
In general, the Fermi surface of transition metals can be distinguished in two areas. The first area is characterized by the $s-$electrons which can be regarded as highly mobile carriers with non-localized wavefunctions and high Fermi velocity, $\upsilon_F$. 
In contrast, the second area is made by electrons which are relatively localized in partially occupied $d-$like states  with a large effective mass and low $\upsilon_F$. 
Thus, it is expected that, the $d-$like states dominate $N$ and, consequently,  $P_0$, while 
the ``{\it light}"  $s-$electrons contribute substantially to the averaged values of Fermi velocity, $\left\langle{\upsilon}_F\right\rangle_\sigma$. As a result,  $P_1$ and $P_2$ may change dramatically if such a contribution in one spin channel is much larger than that in the opposite one (e.g. if $\left\langle{\upsilon}_F\right\rangle_\uparrow \gg \left\langle{\upsilon}_F\right\rangle_\downarrow$).

\begin{figure}[t]
  \begin{center}
\rotatebox{-0}{\includegraphics[width=3.3 in]{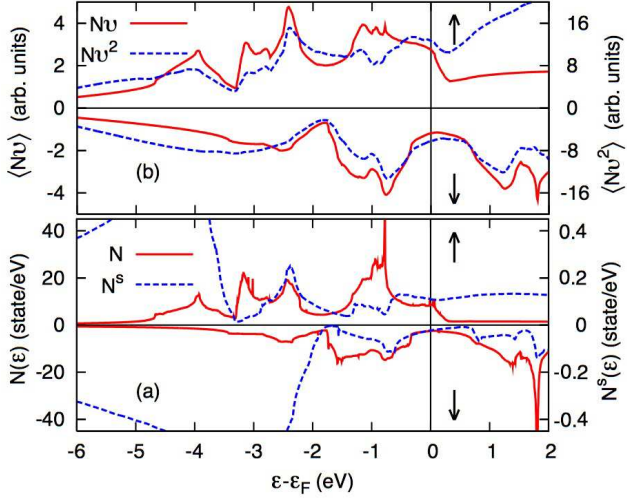}} 
  \end{center}
  \caption{(Color online) Spin polarized (a) total and $s-$projected DOS, $N$ and $N^s$, respectively, as well as (b) Sharvin conductance $\left\langle{N\upsilon}\right\rangle$ and plasma frequency $\left\langle{N\upsilon^2}\right\rangle$ in Fe. The up- and down-states are shown by arrow.}
  \label{fig:Fe}
\end{figure}
\begin{figure}[t]
  \begin{center}
\rotatebox{-0}{\includegraphics[width=3.3 in]{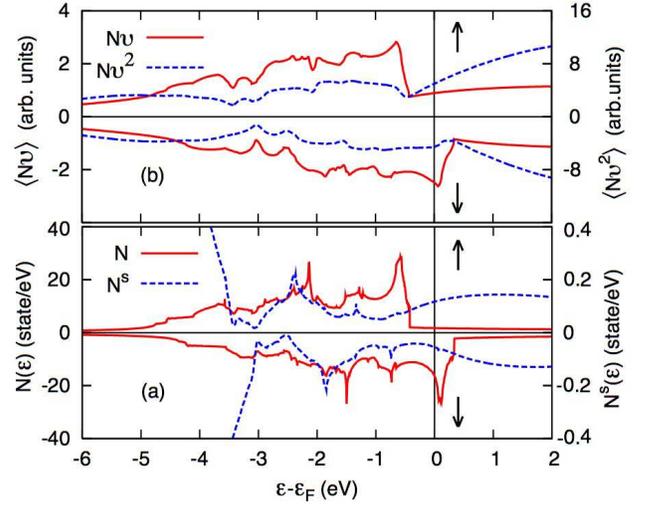}} 
  \end{center}
  \caption{(Color online) Spin polarized (a) total and $s-$projected DOS, $N$ and $N^s$, respectively, as well as (b) Sharvin conductance $\left\langle{N\upsilon}\right\rangle$ and plasma frequency $\left\langle{N\upsilon^2}\right\rangle$ in Ni. }
  \label{fig:Ni}
\end{figure}

For a better understanding, we illustrate in figures~\ref{fig:Fe} and  ~\ref{fig:Ni}, both spin polarized total and $s-$projected DOS,  $N_\sigma$ and  $N^s_\sigma$, respectively, as well as the Sharvin conductance, $\left\langle{N\upsilon}\right\rangle_\sigma$, and the plasma frequency, $\left\langle{N\upsilon^2}\right\rangle_\sigma$ for  Fe and Ni , respectively. In Fe, it is evident that,  both $N_\sigma$ and $N^s_\sigma$ are relatively larger than their corresponding spin-down values. Thus, the obtained spin polarizations  in all limits are positive ($P_n>0$).
Additionally, figure~\ref{fig:Fe}-a indicates a large $s-d$ hybridization in both spin channels, implying that the averaged Fermi velocities, $\left\langle{\upsilon}_F\right\rangle_\uparrow$ and $\left\langle{\upsilon}_F\right\rangle_\downarrow$ should not be very different. Hence, the weighting factors, $\upsilon_{F}$ and $\upsilon^2_{F}$, in eq.~\ref{ballusive} do not considerably change $P_1$ and $P_2$ in comparison with $P_0$, as noted in table~\ref{tab:spinp} and can be seen in figure~\ref{fig:Fe}-b.

 On the other hand, Ni behaves differently. In this case, the spin-up $d-$bands are fully occupied and deeply located below $\varepsilon_F$, so that there is almost no spin-up $d-$state at $\varepsilon_F$, whereas, the spin-down $d-$bands  are partially occupied and have a significant contribution to the $N_\downarrow$ (see figure~\ref{fig:Ni}-a). This leads to a large negative polarization of DOS, ($P_0\ll0$). 
Here again $N^s_\uparrow>N^s_\downarrow$. However, unlike in Fe,  the difference between the averaged Fermi velocities    is so considerable that according to table~\ref{tab:spinp}, $\left\langle{\upsilon}_F\right\rangle_\uparrow$  is almost four times larger than $\left\langle{\upsilon}_F\right\rangle_\downarrow$. This can be attributed to the fact 
that in the absence of ``{\it heavy}" $d-$electrons, $N_\uparrow$ is mainly $s-$like and, hence, the transport carriers are expected to be very mobile with high $\upsilon_F$, while in the spin-down channel, the carriers are dominantly $d-$like and almost immobile in transport.
In the other word, although the number of electrons at $\varepsilon_F$ in spin-up channel is much smaller than that in spin-down channel ($N_\uparrow\ll N_\downarrow$) but they have more tendency to contribute to the transport ($\left\langle{\upsilon}_F\right\rangle_\uparrow\gg\left\langle{\upsilon}_F\right\rangle_\downarrow$).
 Consequently, such an effect results in a significant change in $P_1$ and $P_2$ values (see figure~\ref{fig:Ni}-b). Due to the same reason, the sign reversal in $P_2$ is also expected.

 Having confirmed the accuracy of our method,  we next investigate the transport properties of  CeMnNi$_4$. Both experiment~\cite{singh2006} and theory~\cite{murugan2006,voloshina2006,mazin2006} indicate that, this  intermetallic compound is a soft ferromagnet with a large magnetic moment ($\sim4.9\mu_B$/Mn) and reasonably high curie temperature ($\sim150K$).
 However, there is a significant difference in the experimental and theoretical values of $P$.
  PCAR measurements by Singh et al.~\cite{singh2006}  yield a relatively large spin polarization, $P_C=66\%$, whereas the previous first-principles calculations, using both LSDA and LSDA+U methods, have estimated much lower values for $P$, varying from $-21\%$  to $3\%$~\cite{voloshina2006,mazin2006}.
  Additionally, our calculations on this particular material reveal a metallic nature with a very low degree of spin polarization.
According to table~\ref{tab:spinp}, both $P_1$ $(-5\%)$ and $P_2$ $(10\%)$ are extremely far from the experimental data.  This implies that, it is not possible to realize, through the calculations,  whether the $P_C$ measurements have been carried out in ballistic or diffusive regimes.  However, It may also imply that such a high degree of spin polarization is unlikely to be achieved in pure CeMnNi$_4$.
\begin{figure}[t]
  \begin{center}
\rotatebox{-0}{\includegraphics[width=3.3 in]{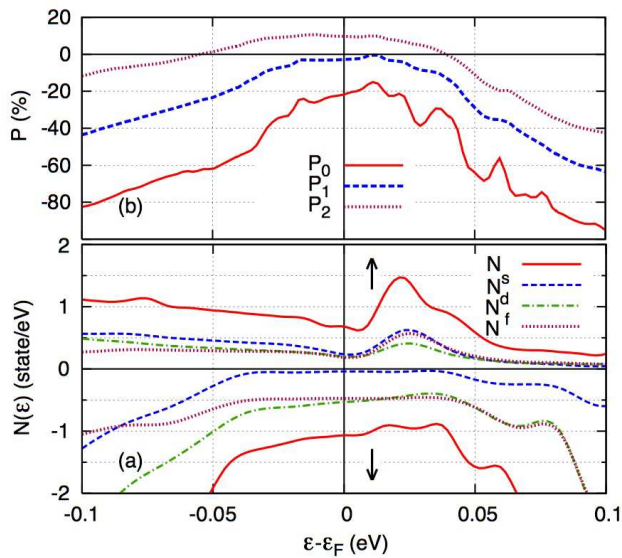}} 
  \end{center}
  \caption {(Color online) Comparison of (a) spin polarized  total, $s-$, $d-$ and $f-$projected DOS, $N$, $N^s$ (scaled by 10), $N^d$ and $N^f$, respectively, and (b) calculated $P_0$, $P_1$ and $P_2$ in CeMnNi$_4$. }
  \label{fig:dsp}
\end{figure}


From the electronic structure point of view, the smallness of $P_n$ values is due to an insignificant difference between  $N_\uparrow$ and $N_\downarrow$ as well as $\left\langle{\upsilon}_F\right\rangle_\uparrow$ and $\left\langle{\upsilon}_F\right\rangle_\downarrow$. To elucidate this situation, we have shown both total and partial DOS of CeMnNi$_4$ around $\varepsilon_F$  in Figure~\ref{fig:dsp}. The figure clearly indicates that  $N_\uparrow$ is slightly smaller than  $N_\downarrow$ and, hence, the calculated $P_0$ turns out to be a small negative value. Also, it is evident from the corresponding partial DOS, $N^d$ and $N^f$, that the $d-$ and $f-$like states dominate $N$ in both spin channels. Consequently, most of the electrons available at $\varepsilon_F$, either $d-$ or $f-$ones, are ``{\it heavy}" and can not effectively contribute to the spin-induced transport.  
Thus, the averaged Fermi velocities $\left\langle{\upsilon}_F\right\rangle_\sigma$ are expected to be very small and close to each other, as denoted in table~\ref{tab:spinp}. Interestingly, our calculations   reveal that both $\left\langle{\upsilon}_F\right\rangle_\uparrow $ and  $\left\langle{\upsilon}_F\right\rangle_\downarrow$ are even relatively (considerably) lower  than the corresponding values in Fe (Ni). The effects, described above, lead to this conclusion that the pure CeMnNi$_4$ is more likely a {\it semi}-metal with poor spin induced transport properties. That is why the calculated $P_0$, $P_1$ and $P_2$ are so small and somehow negligible. Here, it is worth mentioning that, the slight difference between $\left\langle{\upsilon}_F\right\rangle_\uparrow$ and $\left\langle{\upsilon}_F\right\rangle_\downarrow$ in CeMnNi$_4$ are due to the fact that the contribution of $s-$like states to $N_\uparrow$ is slightly larger than that to $N_\downarrow$ (see figure~\ref{fig:dsp}-a). 

It is important to point out that, although the obtained $P_n$ values at $\varepsilon_F$ are substantially smaller than the experimental data, but for changes of the Fermi level by approximately 0.1 eV, values close to experimentally ones for $P_0$ and $P_1$ are reached (see figure~\ref{fig:dsp}-b). Experimentally, this situation can occur if there is a small stoichiometric variation in the sample.
For example, as noted in Ref.~\onlinecite{voloshina2006}, if during the process of preparation, small amount of Mn (e. g. $\sim 5\%$) is replaced by Ni one may expect a considerable change in the electronic structure and hence the  transport properties of the chemically disordered structure. 
In the other word, Since the exact experimental conditions of synthesis of CeMnNi$_4$ is still not completely understood, it might be possible to attribute the measured $P_C$ value to a sample, containing small amount of impurities, defects or chemical disorders.  
Thus, we suggest further studies on the role of stoichiometric variations on the transport properties of CeMnNi$_4$. 

\section{conclusion}
A general and accurate scheme based on tetrahedron method for determining the degree of spin polarization in ferromagnets has been presented.  Our approach has successfully shown to yield highly accurate spin polarization results both in ballistic and diffusive regimes, with the least number of {\it k-}points and the modest computational efforts. The benchmark calculations on bulk Fe and Ni have proven the efficiency of the method in reproducing the experimental data in different regimes. It turned out that the spin polarization in a ferromagnet is dependent on both the number and the type of transport carriers in each spin channel. That is, the more the contribution of $s-$electrons to the density of states in one spin channel, the higher is the degree of spin polarization in both ballistic and diffusive regimes. Finally, in agreement with previous first-principles calculations, the obtained spin polarization in  CeMnNi$_4$ indicated a remarkable difference with the experimental data. The reason was attributed to the strong $d-f$ hybridization in both spin channels, leading to a reduction in contribution of $s-$electrons to the spin induced current. It was suggested that the possible presence of  off-stoichiometery in the sample might be responsible for such a discrepancy.  

\section{acknowledgment}
The authors gratefully acknowledge the Center for Computational Materials Science at the Institute for Materials Research for allocation on the Hitachi SR8000 supercomputer system. We would like to thank Dr. P. Raychauhuri for helpful discussions.


\end{document}